\def\bar{\overline}

\def\a{\alpha}
\def\b{\beta}

\def\th{\theta}

\def\bar{\overline}

\def\eV{{\rm eV}}
\documentstyle [12pt,epsf]{article}
\begin{document}
\baselineskip=22 pt
\setcounter{page}{1}
\thispagestyle{empty}
\topskip 0.5  cm
\begin{flushright}
\begin{tabular}{c c}
& {\normalsize  NSF-KITP-03-82}\\
&  September 2003
\end{tabular}
\end{flushright}
\vspace{1 cm}
\centerline{\LARGE \bf    Prediction of $\bf U_{e3}$}
\vskip 0.5 cm
\centerline{\LARGE \bf  in Neutrino Mass Matrix with Two Zeros}
\vskip 1.5 cm
\centerline{{\large \bf Mizue Honda}
 \renewcommand{\thefootnote}{\fnsymbol{footnote}}
\footnote[1]{E-mail address:  mizue@muse.sc.niigata-u.ac.jp},
\qquad{\large \bf Satoru Kaneko}
\renewcommand{\thefootnote}{\fnsymbol{footnote}}
\footnote[2]{E-mail address: kaneko@muse.sc.niigata-u.ac.jp}}
\vskip 0.3 cm
\centerline{\large\bf and}
\vskip 0.3 cm
\centerline{{\large \bf Morimitsu Tanimoto}
\renewcommand{\thefootnote}{\fnsymbol{footnote}}
\footnote[3]{E-mail address: tanimoto@muse.sc.niigata-u.ac.jp}
 }
\vskip 0.8 cm
 \centerline{ \it{Department of Physics, Niigata University, 
 Ikarashi 2-8050, 950-2181 Niigata, JAPAN}}
\vskip 2 cm
\centerline{\bf ABSTRACT}\par
\vskip 1 cm
We  have discussed  predictions of  $|U_{e3}|$ and $J_{CP}$ in the 
framework of the neutrino mass matrix with two zeros.
In the case of the best fit values  of $\tan^2\theta_{12}$,
$\tan^2\theta_{23}$, $\Delta m^2_{\rm sun}$ and $\Delta m_{\rm atm}^2$, 
the prediction of  $|U_{e3}|$  is $0.11\sim 0.14$.
The lower bound of $|U_{e3}|$ is $0.05$, which  depends on
  $\tan\theta_{12}$ and $\tan\theta_{23}$. We  have investigated
  the stability of these predictions taking account 
of small corrections to zeros, which may come from radiative corrections or 
off-diagonal elements of the charged lepton mass matrix. 
The lower bound of $|U_{e3}|$ comes down considerably 
 due to the small corrections to zeros.

\newpage
\topskip 0. cm
\section{Introduction}
 In recent  years empirical understanding of the mass and mixing of 
 neutrinos  have been  advanced \cite{SKam,SKamsolar,SNO}.
The KamLAND experiment selected the neutrino
mixing solution that is responsible for the solar neutrino problem
nearly uniquely \cite{KamLAND},   only large mixing angle solution.
We have now good understanding concerning the neutrino
mass difference squared ($\Delta m^2_{\rm atm}$, $\Delta m^2_{\rm sun}$) and 
neutrino flavor mixings ($\sin^2 2\theta_{\rm atm}$,
 $\tan^2 \theta_{\rm sun}$) \cite{Lisi}. A  constraint has also
been placed on the mixing from the reactor experiment of CHOOZ \cite{CHOOZ}.

The texture zeros of the neutrino mass matrix have been discussed
 to explain these  neutrino masses and mixings  \cite{Fu,AK,Kang}.
  Recently, Frampton, Glashow and Marfatia \cite{Fram} found 
acceptable textures of the  neutrino mass matrix with two independent
vanishing entries in the basis of the diagonal charged lepton mass matrix.
 The  KamLAND result has stimulated the phenomenological analyses
of the texture zeros
\cite{Xing1,Xing2,Barbieri,Obara}.
These results favor texture zeros of the neutrino mass matrix
phenomenologically.

 However, there are theoretical problems.
The first one is the effect of radiative corrections.
A specific texture of the lepton mass matrix is  not  preserved to
all orders.  For example,  non-zero components may evolve in zero-entries of 
the mass matrix at the low energy scale
 due to radiative corrections even if the zero texture is realized at the high
energy scale.
The second one is the choice of the flavor basis.
In the model with some flavor symmetry, zeros of the neutrino mass matrix are
 given  while the charged lepton mass matrix has off-diagonal components
\footnote{There are some models in which zeros of the neutrino mass matrix are
 realized in the diagonal  basis of the charged lepton mass matrix 
\cite{Zee,FramY,our}.}. 
Then, zeros of the neutrino mass matrix are polluted after diagonalizing
  the charged lepton mass matrix.
  Therefore we need to study the stability of
 predictions  of  the texture zeros 
 by  taking account of small corrections to zeros.

In this paper, we present detailed study of the neutrino mixing parameter 
$U_{e3}$ and CP violating quantity $J_{CP}$ \cite{JCP}, 
which are expected to be affected  by the small corrections
to zeros in the neutrino mass matrix.
It is found that the predicted  $U_{e3}$ and $J_{CP}$ 
considerably depend  on these corrections.

Predictions of the texture two zeros are presented in section 2.  
Small corrections to zeros are discussed and the stability of 
predictions are studied in section 3.
Section 4 is devoted to the summary.

\section{Predictions of texture two zeros}
There are  15 textures with two zeros  for the effective neutrino mass matrix 
$M_\nu$, which have five independent parameters. 
The two zero conditions give
\begin{equation}
 (M_\nu)_{ab}=\sum^3_{i=1} U_{ai} U_{bi}\lambda_i = 0 \ , \qquad\qquad
 (M_\nu)_{\a\b}=\sum^3_{i=1} U_{\alpha i} U_{\beta i}\lambda_i = 0 \ ,
\label{cons1} 
\end{equation}
\noindent  where $\lambda_i$ is the i-th eigenvalue
including the  Majorana phases, and indices $(a b)$ and $ (\a  \b)$ denote 
 the  flavor components, respectively.

Solving these equations,  ratios of neutrino masses $m_1$, $m_2$, $m_3$,
 which are absolute values of $\lambda_i$'s, are given
in terms of the neutrino mixing matrix  $U$ \cite{MNS} as follows:
\begin{eqnarray} 
\frac{m_1}{m_3} & = & \left |
\frac{U_{a3} U_{b3} U_{\alpha 2} U_{\beta 2} - U_{a2} U_{b2} U_{\alpha 3}
U_{\beta 3}}{U_{a2} U_{b2} U_{\alpha 1} U_{\beta 1} - U_{a1} U_{b1}
U_{\alpha 2} U_{\beta 2}} \right | \ ,
\nonumber \\ 
\frac{m_2}{m_3} & = & \left |
\frac{U_{a1} U_{b1} U_{\alpha 3} U_{\beta 3} - U_{a3} U_{b3} U_{\alpha 1}
U_{\beta 1}}{U_{a2} U_{b2} U_{\alpha 1} U_{\beta 1} - U_{a1} U_{b1}
U_{\alpha 2} U_{\beta 2}} \right |  \ .
\end{eqnarray}
\noindent Then, one can test textures in the ratio $R_\nu$,
\begin{eqnarray}
R_\nu \equiv \left | \frac{m^2_2 - m^2_1}
{m^2_3 - m^2_2} \right | \approx \frac{\Delta m^2_{\rm sun}}
{\Delta m^2_{\rm atm}} \  ,
\label{Rnu}
\end{eqnarray}
\noindent which has been  given by the experimental data.
 The ratio $R_\nu$ is given only in terms of four parameters 
 (three mixing angles and one phase) in 
\begin{eqnarray}
 U = \left (\matrix{ c_{13} c_{12} & c_{13} s_{12} &  s_{13} e^{-i \delta}\cr 
  -c_{23}s_{12}-s_{23}s_{13}c_{12}e^{i \delta}
 & c_{23}c_{12}-s_{23}s_{13}s_{12}e^{i \delta} &   s_{23}c_{13} \cr
  s_{23}s_{12}-c_{23}s_{13}c_{12}e^{i \delta} 
& -s_{23}c_{12}-c_{23}s_{13}s_{12}e^{i \delta} & c_{23}c_{13} \cr}\right )\ ,
\end{eqnarray}
\noindent where $c_{ij}$ and  $s_{ij}$ denote
 $\cos \theta_{ij}$ and $\sin \theta_{ij}$, respectively.

 Seven acceptable textures with two independent 
zeros were found for the  neutrino mass matrix \cite{Fram}, 
and they have been studied in detail
\cite{Xing2,Barbieri}
\footnote{Additional two textures may be allowed marginally by current data
 as shown in ref.\cite{Xing2}.}.
Among them, the textures $A_1$ and $A_2$ of ref.\cite{Fram}, which  
correspond to the hierarchical neutrino mass spectrum, are strongly favored 
 by the recent phenomenological analyses \cite{Xing1,Xing2,Barbieri}.
  Therefore,  we study these two  textures in this paper.

In the texture $A_1$,
 which has two zeros as  $(M_\nu)_{ee}=0$ and  $(M_\nu)_{e\mu}=0$, 
the mass ratios are given as
\begin{eqnarray}
\frac{m_1}{m_3} & = &
 \left | \frac{s_{13}}{c^2_{13}} \left (\frac{s_{12} s_{23}}{c_{12} c_{23}}-
 s_{13}  e^{-i\delta}\right ) \right |\ ,
\nonumber \\
\frac{m_2}{m_3} & = &
 \left |\frac{s_{13}}{c^2_{13}} \left ( \frac{c_{12} s_{23}}{s_{12} c_{23}} +
 s_{13}  e^{-i\delta}\right )\right | \ .
\label{mA1}
\end{eqnarray}
\noindent In  the texture $A_2$,
 which has two zeros as  $(M_\nu)_{ee}=0$ and  $(M_\nu)_{e\tau}=0$, 
the mass ratios are given as
\begin{eqnarray}
\frac{m_1}{m_3} & = &
 \left | \frac{s_{13}}{c^2_{13}} \left (\frac{s_{12} c_{23}}{c_{12} s_{23}}+
 s_{13}  e^{-i\delta}\right ) \right | \ ,
\nonumber \\
\frac{m_2}{m_3} & = &
 \left |\frac{s_{13}}{c^2_{13}} \left ( \frac{c_{12} c_{23}}{s_{12} s_{23}} -
 s_{13}  e^{-i\delta}\right )\right | \ .
\label{mA2}
\end{eqnarray}
\noindent
It is remarked that  the mass ratios of the
texture $A_2$ are given  exactly by replacing 
$\tan\theta_{23}$ in $A_1$ with  $-\cot\theta_{23}$.
 
If $\theta_{12}$, $\theta_{23}$,  $\theta_{13}$ and $\delta$ are fixed,
 we can predict $R_\nu$ of eq.(\ref{Rnu}), which  is  compared with 
the experimental value ${\Delta m^2_{\rm sun}}/{\Delta m^2_{\rm atm}}$.
Taking account of the following data with $90\%$ C.L. \cite{Lisi},
\begin{eqnarray}
 \sin^2 2\th_{\rm atm} \geq 0.92\ , \qquad
 &&\Delta m^2_{\rm atm}=  (1.5\sim 3.9)\times  10^{-3} \eV^2\ , \nonumber \\
 \tan^2 \th_{\rm sun}=0.33\sim 0.67\ ,  \qquad
 &&\Delta m_{\rm sun}^2= (6\sim 8.5)\times 10^{-5}\eV^2\ ,
\label{Data}
\end{eqnarray}
\noindent with  $\sin \th_{\scriptscriptstyle \rm CHOOZ} \leq 0.2$, 
we predict $R_\nu$.
In Fig.1,  we present the scatter plot of  the predicted $R_\nu$ versus 
$\sin\theta_{13}$,
in which $\delta$ is taken in the whole range $-\pi \sim \pi$ 
for the texture $A_1$.
 The parameters are taken in the following ranges of 
 $\theta_{12}=30^\circ\sim 39^\circ$, $\theta_{23}=37^\circ\sim 53^\circ$,
 $\theta_{13}=1^\circ\sim 12^\circ$ and $\delta=-\pi\sim\pi$.
It is found that 
many predicted values of $R_\nu$ lie outside the experimental allowed region. 
This result means  that 
there is a strong constraint for the parameter $\theta_{13}$.
We get $\sin\theta_{13}\geq 0.05$ from
the experimental value  of $R_\nu$ as seen in Fig.1. 

In order to present the allowed  region of $\sin\theta_{13}$
for the texture $A_1$,
 we show the scatter plot  of $\sin\theta_{13}$ versus
$\tan^2\theta_{12}$ and $\tan^2\theta_{23}$ in Fig.2 and Fig.3, respectively.
If we take the best fit values  of $\tan^2\theta_{12}=0.42$,  
$\tan^2\theta_{23}=1$, $\Delta m_{\rm sun}^2= 7.3\times 10^{-5}\eV^2$ and
$\Delta m^2_{\rm atm}=2.5\times 10^{-3}\eV^2$,
 the prediction of  $\sin\theta_{13}$  is
$0.11\sim 0.14$,  where the phase $\delta$  is taken 
in the whole range $-\pi \sim \pi$. 
 In Fig.4, we show $J_{CP}$ versus
 $\sin\theta_{13}$ for  the texture $A_1$.
 Since  $J_{CP}$ is proportional to $\sin\theta_{13}$ and $\sin\delta$,
 we also show  the allowed region of $\delta$  versus $\sin\theta_{13}$
in Fig.5.  It is found that  $\delta$ is allowed in the whole range of
 $-\pi\sim \pi$.

 We do not show the numerical results in the texture $A_2$
 because  those are obtained only by replacing 
$\tan\theta_{23}$ in $A_1$ with  $-\cot\theta_{23}$.
 
The allowed regions in Fig.2 and Fig.3 are  quantitatively  understandable
in the following approximate relations:
\begin{eqnarray}
 |U_{e3}|\equiv\sin\theta_{13}\simeq
  \frac{1}{2}\tan 2\theta_{12} \ \cot \theta_{23}
\sqrt{R_\nu \cos 2\theta_{12}} \ ,
\end{eqnarray}
\noindent for the texture $A_1$, and 
\begin{eqnarray}
 |U_{e3}|\equiv\sin\theta_{13}\simeq \frac{1}{2}\tan 2\theta_{12} \ \tan \theta_{23}
\sqrt{R_\nu \cos 2\theta_{12}} \ ,
\end{eqnarray}
\noindent for the texture $A_2$, respectively,
where the phase $\delta$  is neglected because it is a next leading term.
 As $\tan\theta_{12}$ increases, the lower bound of $|U_{e3}|$ increases,
and as  $\tan\theta_{23}$ decreases, it  increases.
It is found in Fig.2 that the lower bound  $|U_{e3}|=0.05$ is given
 in the case of   the smallest $\tan^2\theta_{12}$, while
 $|U_{e3}|=0.08$  is given in the largest  $\tan^2\theta_{12}$.
On the other hand, as seen in Fig.3, the lower bound $|U_{e3}|=0.05$ is given 
in the largest $\tan^2\theta_{23}$, while $|U_{e3}|=0.08$ is given
 in  the smallest  $\tan^2\theta_{23}$.
The upper bound of $|J_{CP}|$ is $0.05$, but $J_{CP}=0$ is still allowed
in Fig.4.
These predicted regions will be reduced in the future since error bars of 
experimental data in eq.(\ref{Data}) will be reduced, especially, 
KamLAND is expected to determine
 $\Delta m^2_{12}$ precisely. 

\section{Stability of  predicted  $\bf U_{e3}$}
 Above  predictions are important ones in  the texture zeros.
The relative magnitude of each entry of the neutrino mass matrix is roughly
given as follows: 
\begin{eqnarray}
M_\nu \sim \left ( \matrix{
{ 0} & {0} & \lambda \cr {0} & 1 & 1 \cr \lambda  & 1 & 1 \cr} \right )
\  \ \ {\rm for}\  A_1 \ , \qquad
\qquad 
\left (\matrix{ 0 & \lambda & 0\cr\lambda  & 1 & 1 \cr 0 & 1 & 1 \cr} \right )
\  \ \ {\rm for} \ A_2 \ ,
\label{A12}
\end{eqnarray}
\noindent
where  $\lambda\simeq 0.2$.   
However,  these  texture zeros  are not  preserved to all orders. 
 Even if zero-entries of the mass matrix are given at the high energy scale,
 non-zero components may evolve instead of zeros at the low energy scale
  due to  radiative corrections.
Those magnitudes depend on unspecified interactions from which lepton masses
are generated. 
Moreover,  zeros  of the neutrino mass matrix are
given  while the charged lepton mass matrix has off-diagonal components
in the model with  some flavor symmetry. 
 Then, zeros are not  realized in the diagonal basis of the 
charged lepton mass matrix.
In other words, zeros of the neutrino mass matrix is  polluted by 
the small off-diagonal elements of the charged lepton mass matrix.

Therefore, one need the careful study of stability of the prediction for 
 $|U_{e3}|$ and $J_{CP}$ because these are  small quantities.
In order to see the effect of the small non-zero components,
the conditions of zeros in  eq.(\ref{cons1}) are modified.
The two  conditions   turn to
\begin{equation}
 (M_\nu)_{ab}=\sum^3_{i=1} U_{ai} U_{bi}\lambda_i = \epsilon \ , \qquad\qquad
 (M_\nu)_{\a\b}=\sum^3_{i=1} U_{\alpha i} U_{\beta i}\lambda_i = \omega \ ,
\label{cons2} 
\end{equation}
\noindent
where $\epsilon$ and $\omega$ are arbitrary parameters with the mass unit,
 which are
much smaller than other non-zero components of the mass matrix.  
These parameters are supposed to be real for simplicity.
We get
\begin{eqnarray} 
\frac{m_1}{m_3} & = & \left |
\frac{U_{13} U_{13} U_{12} U_{22} - U_{12} U_{12} U_{13}U_{23}
-U_{12} U_{22}\bar\epsilon + U_{12} U_{12}\bar\omega}
 {U_{12} U_{12} U_{11} U_{21} - U_{11} U_{11} U_{12} U_{22}} \right | \ ,
\nonumber \\ 
\frac{m_2}{m_3} & = & \left |
\frac{U_{11} U_{11} U_{13} U_{23} - U_{13} U_{13} U_{11}U_{21}
+U_{11} U_{21}\bar\epsilon - U_{11} U_{11}\bar\omega}
{U_{12} U_{12} U_{11} U_{21} - U_{11} U_{11}U_{12} U_{22}} \right |  \ ,
\label{ratio}
\end{eqnarray}
\noindent
 for the texture $A_1$,  and 
\begin{eqnarray} 
\frac{m_1}{m_3} & = & \left |
\frac{U_{13} U_{13} U_{12} U_{32} - U_{12} U_{12} U_{13}U_{33}
-U_{12} U_{32}\bar\epsilon + U_{12} U_{12}\bar\omega}
 {U_{12} U_{12} U_{11} U_{31} - U_{11} U_{11} U_{12} U_{32}} \right | \ ,
\nonumber \\ 
\frac{m_2}{m_3} & = & \left |
\frac{U_{11} U_{11} U_{13} U_{33} - U_{13} U_{13} U_{11}U_{31}
+U_{11} U_{31}\bar\epsilon - U_{11} U_{11}\bar\omega}
{U_{12} U_{12} U_{11} U_{31} - U_{11} U_{11}U_{12} U_{32}} \right |  \ ,
\label{ratioA2}
\end{eqnarray}
\noindent  for the texture $A_2$, 
where $\bar\epsilon$ and $\bar\omega$ are normalized ones as
$\bar\epsilon=\epsilon/\lambda_3$ and $\bar\omega=\omega/\lambda_3$, 
respectively.   We obtain approximately   
\begin{eqnarray}
\frac{m_1}{m_3} & \simeq &  \left | t_{12} t_{23} s_{13} e^{-i \delta}- 
\frac{t_{12}}{c_{23}}\bar\omega + \bar\epsilon \right | \ ,
\nonumber \\
\frac{m_2}{m_3} &\simeq &\left | \frac{1}{t_{12}} t_{23} s_{13} e^{-i \delta}
 - \frac{1}{t_{12}c_{23}}\bar\omega - \bar\epsilon \right |\ ,
\label{mmA1}
 \end{eqnarray}
\noindent  for the texture $A_1$, and
\begin{eqnarray}
\frac{m_1}{m_3} &\simeq &\left | -t_{12}\frac{1}{t_{23}}s_{13} e^{-i \delta}+
\frac{t_{12}}{s_{23}}\bar\omega +\bar\epsilon \right | \ ,
\nonumber \\
\frac{m_2}{m_3} &\simeq& \left | -\frac{1}{t_{12} t_{23}} s_{13} e^{-i \delta}
 + \frac{1}{t_{12}s_{23}}\bar\omega -\bar\epsilon \right |\ ,
\label{mmA2}
 \end{eqnarray}
\noindent  for the texture $A_2$, where $t_{ij}=\tan\theta_{ij}$.
We present the approximate expression of $|U_{e3}|=\sin\theta_{13}$ 
as follows:
\begin{eqnarray}
 \sin\theta_{13}\simeq \pm \frac{1}{2}\tan 2\theta_{12}\ \cot\theta_{23}
\sqrt{R_\nu \cos 2\theta_{12}} + \frac{\bar\omega \  \cos \delta}
{s_{23}}  +
 \frac{t_{12}}{t_{23}}\frac{\bar\epsilon \ \cos\delta}{1- t^2_{12}}
+O(\bar\omega^2,\ \bar\epsilon^2)\ ,
\label{Ue3corr}
\end{eqnarray}
\noindent
 for the texture $A_1$, and  
\begin{eqnarray}
 \sin\theta_{13}\simeq \pm\frac{1}{2}\tan 2\theta_{12}\ \tan\theta_{23}
\sqrt{R_\nu \cos 2\theta_{12}} + \frac{\bar\omega \  \cos \delta}
{c_{23}}  -
 t_{12}t_{23}\frac{\bar\epsilon \ \cos\delta}{1-t^2_{12}}
+O(\bar\omega^2,\ \bar\epsilon^2)\ ,
\label{Ue3corr2}
\end{eqnarray}
\noindent
 for the texture  $A_2$.
In these equations,  $+(-)$ of the first term in  the right-hand side  
 corresponds to  the case of $\tan\theta_{12}\tan\theta_{23}>0 (<0)$.
It is remarked that the second and third terms in the right-hand side could 
be comparable with the first one in  eq.(\ref{Ue3corr}) and  
eq.(\ref{Ue3corr2}).  

However, the second and third terms in the right-hand side 
 could be partially canceled each other depending on the  sign of
   $\tan\theta_{12}$ and   $\tan\theta_{23}$.
Therefore, the prediction of  $|U_{e3}|$ is  somewhat different
 between  the texture  $A_1$ and  $A_2$.

In order to estimate the effect of  $\bar\omega$ and  $\bar\epsilon$,
  we consider the case in which the charged lepton mass matrix has 
 small off-diagonal components. 
Suppose that the  two zeros 
 in eq.(\ref{A12}) is still preserved for the neutrino sector.
The typical model of the charged lepton 
 is the Georgi-Jarlskog texture \cite{GJ},
in which the charged lepton mass matrix $M_E$ is given as
\begin{equation}
M_E \simeq  \ \left ( \matrix{
 0 &\sqrt{m_e m_\mu} & 0 \cr \sqrt{m_e m_\mu} & m_\mu & \sqrt{m_e m_\tau} \cr
 0  & \sqrt{m_e m_\tau} & m_\tau \cr} \right ) \ ,
\label{GJtex}
\end{equation}
\noindent where each matrix element is written in terms of the
charged lepton masses,  and phases are neglected for simplicity.
This matrix is diagonalized by the unitary matrix $U_E$, 
in which the mixing between the first and second families 
is $\sqrt{\frac{m_e}{m_\mu}}\simeq 0.07$ and 
the mixing between the second  and third families 
is $\sqrt{\frac{m_e}{m_\tau}}\simeq 0.02$.
Since the neutrino mass matrix is still the texture $A_1$ or  
the texture $A_2$ \footnote{The combined model with 
Georgi-Jarlskog texture and the $A_2$ type one is presented 
in ref.\cite{Obara}},
  it turns to $A_1'$ or  $A_2'$  as follows:
\begin{eqnarray}
M_\nu \sim \ \ \left ( \matrix{ \kappa^2 &\kappa &\lambda \cr
\kappa & 1 & 1 \cr \lambda  & 1 & 1 \cr} \right ) \quad {\rm for \ }A_1' \ ,
 \qquad
 \left ( \matrix{ 2\kappa\lambda & \lambda &\kappa \cr
\lambda & 1 & 1 \cr \kappa  & 1 & 1 \cr} \right )  \quad {\rm for \ }A_2' \ ,
\label{A1modi}
\end{eqnarray}
\noindent in the diagonal basis of the charged lepton mass matrix. 
 Therefore, the parameter    $\bar\epsilon$ 
are correlated with $\bar\omega$ such as  $\bar\omega\simeq \kappa/2$
and  $\bar\epsilon\simeq \kappa^2/2$  in the texture $A_1'$, 
and  $\bar\omega\simeq \kappa/2$
and  $\bar\epsilon\simeq  \kappa\lambda$  in the texture $A_2'$.

By using the texture $A_1'$ of the neutrinos in  eq.(\ref{A1modi}),
we show our results of  the allowed region of $\sin\theta_{13}$ versus
$\tan^2\theta_{12}$ and $\tan^2\theta_{23}$ in Fig.6 and Fig.7, respectively,
where $\tan\theta_{12}$ and $\tan\theta_{23}$ are taken to be positive.
In the case of the best fit values of 
 $\tan^2\theta_{12}$, $\tan^2\theta_{23}$, $\Delta m^2_{\rm sun}$ 
and $\Delta m_{\rm atm}^2$, the prediction of $\sin\theta_{13}$ is
$0.075\sim 0.15$,  
which is somewhat wider than  the result of the section 2 due to the
correction $\kappa$.
We have also calculated in the case of 
 $\tan\theta_{12}\tan\theta_{23}<0$.  Predictions are almost same
because of $\bar \omega \gg \bar \epsilon$.
In Fig.8,  we show $J_{CP}$ versus $\sin\theta_{13}$.
 We also show  the allowed region of $\delta$  versus $\sin\theta_{13}$
in Fig.9. 
These results should be compared with the ones in Fig.2 $\sim$ Fig.5.
 It is noticed that the lower bound of $\sin\theta_{13}$ 
considerably comes down due to the correction $\kappa$.
  The small $|U_{e3}|$ of  $8\times 10^{-3}$ is allowed.
  In Fig.6 and Fig.7, we also find  that  
there is an  upper bound on $|U_{e3}|$  for small values of
$\tan^2\theta_{12}$ and large values of $\tan^2\theta_{23}$.  
We have checked numerically that this upper bound appears due to 
the higher order terms, which are  omitted
in eqs.(\ref{mmA1}), (\ref{mmA2}), (\ref{Ue3corr}) and (\ref{Ue3corr2}).
As seen in Fig.9, the phase $\delta$ is not allowed in the whole range 
$-\pi\sim \pi$ if $\sin\theta_{12}\leq 0.09$.

In the texture $A_2'$  of  eq.(\ref{A1modi}),
we show  allowed regions of $\sin\theta_{13}$ versus
$\tan^2\theta_{12}$ and $\tan^2\theta_{23}$ in Fig.10 and Fig.11, 
respectively.
In the case of the best fit values of  $\tan^2\theta_{12}$, 
$\tan^2\theta_{23}$, $\Delta m^2_{\rm sun}$ and $\Delta m_{\rm atm}^2$, 
the prediction of  $\sin\theta_{13}$  is
$0.08\sim 0.195$, which is somewhat different from the result of  $A_1'$.
There is no  upper bound on $|U_{e3}|$ 
in  Fig.10 and Fig.11.  The effect of $\bar\omega$ is canceled partially by 
 the effect of $\bar\epsilon$  as seen in eq.(\ref{Ue3corr2}). 
Therefore, allowed regions  in Fig.10 and Fig.11 are somewhat different from 
ones in Fig.6 and Fig.7. 
We show the result in the case of 
 $\tan\theta_{12}\tan\theta_{23}<0$ in Fig.12 and Fig.13.  
 The lower bound of $\sin\theta_{13}$ 
  comes down to zero, which is contrasted with the result in Fig.10 and Fig.11 because there is no cancellation between the effect of $\bar\omega$ and
  $\bar\epsilon$  as seen in eq.(\ref{Ue3corr2}). 
 We omit the figure of  $J_{CP}$ because it is almost same as the result
 in the case of  $A_1'$.

This result depends on a specific model.
In order to complete the study of the stability, we also consider the case 
that two corrections  $\bar\epsilon$ and $\bar\omega$ are independent 
 each other, which may be the case if the deviation from zero
arises from radiative corrections.
For the texture $A_1'$,  we show the allowed  region
 of $\sin\theta_{13}$ versus $2\bar\omega$ in Fig.14,
where $\bar \epsilon=0$ is taken.
 The gray region is allowed by the experimental data for the texture $A_1'$.
The allowed region of the texture $A_2'$ is almost same as the one of $A_1'$. 
   In Fig.14, the narrow deep gray region is only added in the case of 
 the texture $A_2'$.
 We also show the allowed region of $\sin\theta_{13}$ versus $\bar\epsilon$ 
 in Fig.15, in which $\bar\omega=0$ is taken.
There is no difference between the allowed region of both $A_1'$ and $A_2'$. 
 
  The predicted lower bound of $|U_{e3}|$ is sensitive to  $\bar \omega$.
 If the correction $\bar\omega$ is as large as  $0.02$, the lower bound 
of $|U_{e3}|$ comes down to $0.025$.
If  $\bar\omega$ is larger than $0.04$, the predictability of 
  $|U_{e3}|$ is lost since  $|U_{e3}|=0$ is allowed.
 On the other hand,  the lower bound  is rather insensitive to $\bar\epsilon$.
Even if the correction $\bar\epsilon$ grows up to be  $0.1$,  
 the lower bound of $|U_{e3}|$ is still  $0.02$.
It is concluded that the magnitude of the deviation from the zero 
in the (1,2)/(1,3) element
for the texture $A_1$/$A_2$ is important to predict  $|U_{e3}|$.

\section{Summary}
 We summarize our results as follows.
We have studied  $|U_{e3}|$ and $J_{CP}$ in the textures $A_1$ and
$A_2$ of the neutrino mass matrix with two zeros.   
The lower bound of $|U_{e3}|$
 is $0.05$, which considerably depends on  $\tan^2\theta_{12}$ and 
$\tan^2\theta_{23}$.
 We have investigated the stability of these predictions 
by taking account of small corrections, which may come from radiative 
corrections or off-diagonal elements of the charged lepton mass matrix.  
The lower bound of $|U_{e3}|$ comes down significantly
in the case of  $\bar\omega\gg 0.01$, while $\bar\epsilon$  
is rather insensitive to  $|U_{e3}|$. Therefore the corrections 
to the texture zeros in the (1,2)/(1,3) element for the texture
 $A_1$/$A_2$ should be carefully taken to predict  $|U_{e3}|$.

In the case of  best fit values  of  $\tan^2\theta_{12}$, 
$\tan^2\theta_{23}$, $\Delta m^2_{\rm sun}$ and $\Delta m_{\rm atm}^2$, 
 the prediction of  $|U_{e3}|$  is $0.075\sim 0.195$
even if the corrections are taken into account.
 This  prediction  is good news for the near future experiments.
 The measurement of  $|U_{e3}|$ will be an important  test of the texture
zeros  in the future.

\vskip 0.5  cm
 
 We would like to thank  M. Katsumata, H. Nakano and H. So
  for helpful discussions.
 We are also grateful to KITP in UCSB, where the paper was completed,
 for its warm hospitality.
 This research was supported in part by the National Science
Foundation in USA under Grant No. PHY99-07949 and the Grant-in-Aid for Science Research, Ministry of Education, Science and Culture, Japan(No.12047220).

\newpage

\newpage
\begin{figure}
\epsfxsize=14.0 cm
\centerline{\epsfbox{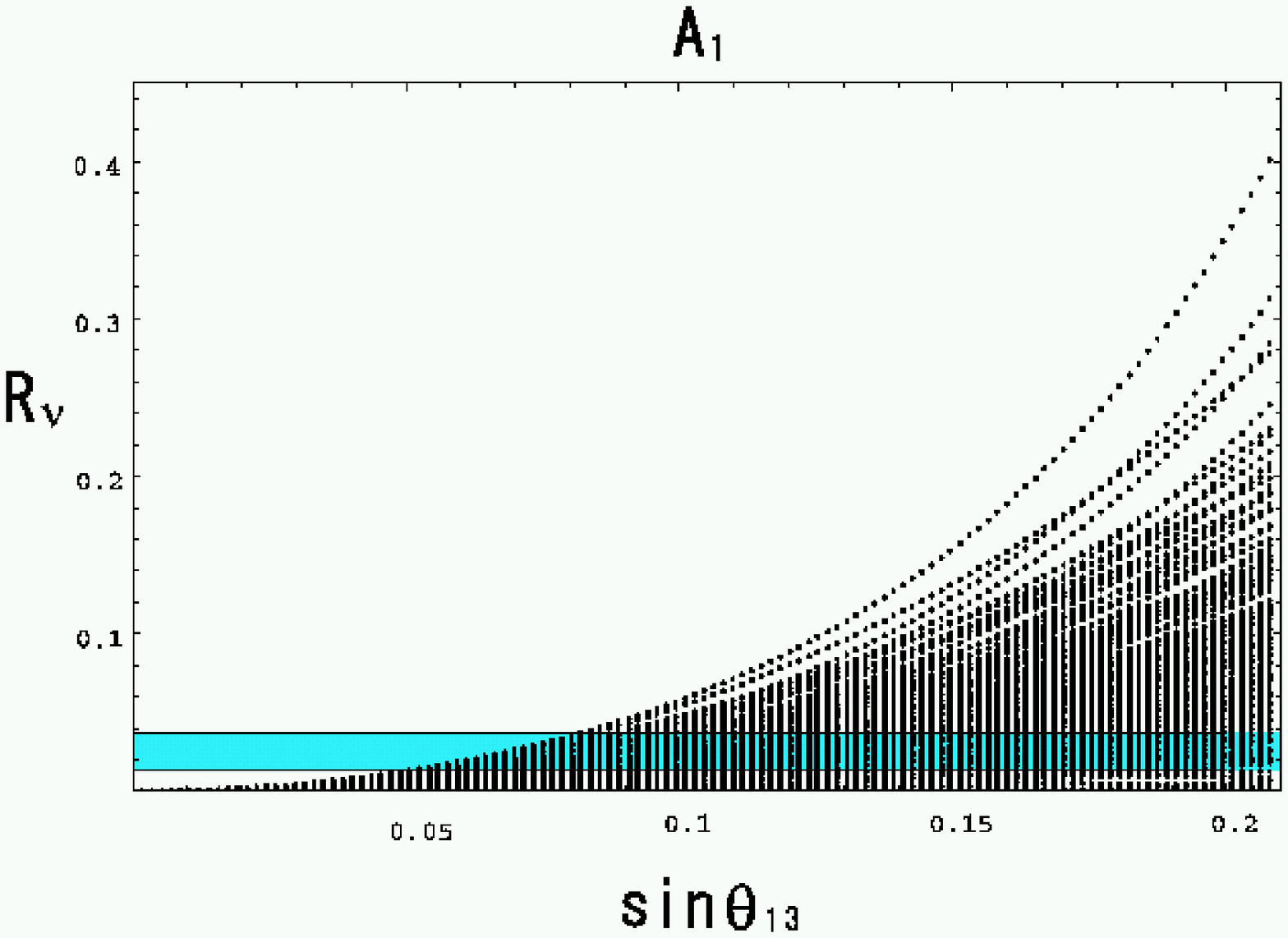}}
\caption{Scatter plot of $R_\nu$  versus $\sin\theta_{13}$ for $A_1$.
 The unknown phase $\delta$ is taken in the whole range $-\pi\sim \pi$.
 The gray horizontal band is the experimental allowed region.}
\end{figure}
\begin{figure}
\epsfxsize=14.0 cm
\centerline{\epsfbox{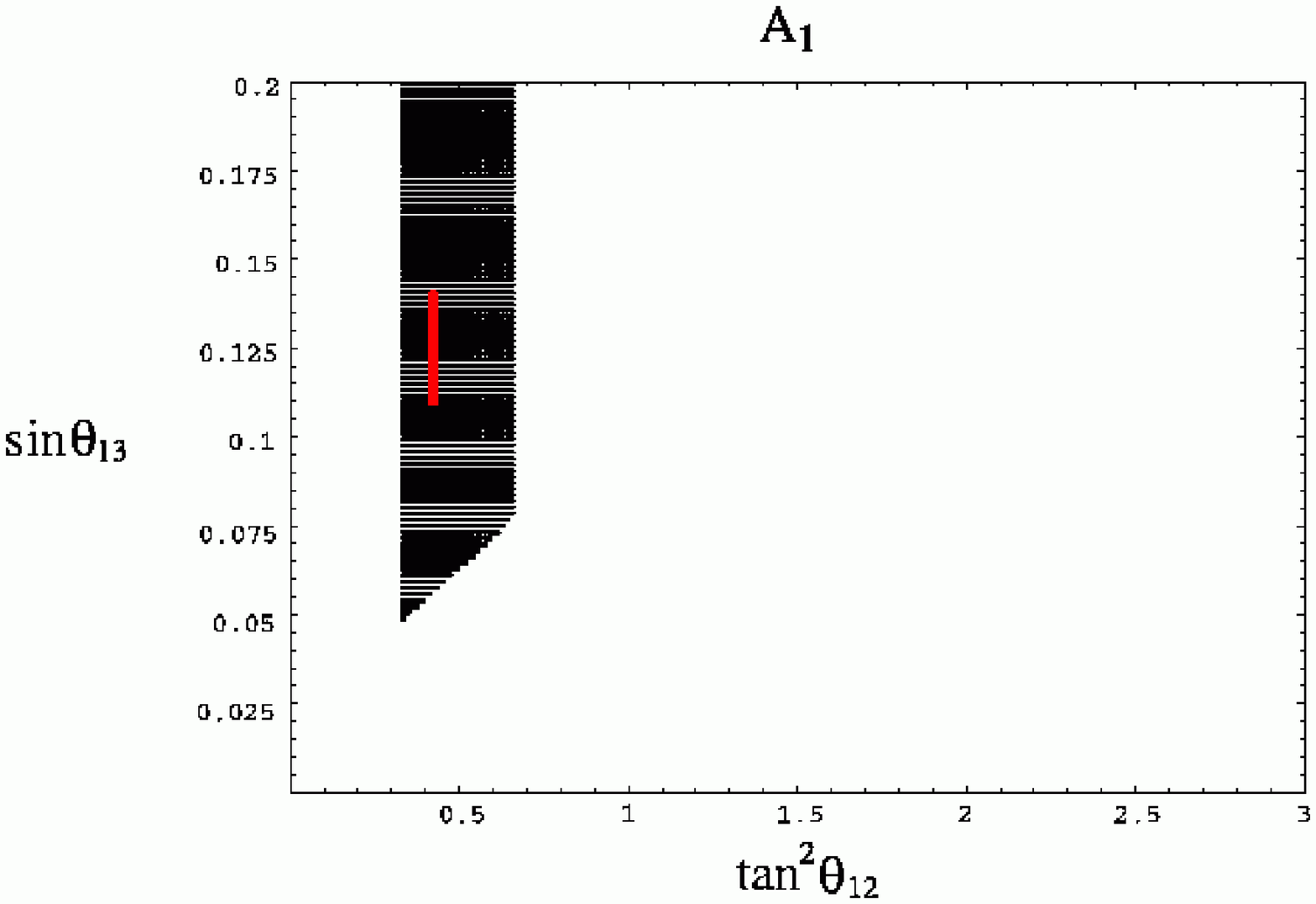}}
\caption{Scatter plot of $\sin\theta_{13}$ versus $\tan^2\theta_{12}$
for $A_1$. The  best fit is shown by a red line.}
\end{figure}
\begin{figure}
\epsfxsize=14.0 cm
\centerline{\epsfbox{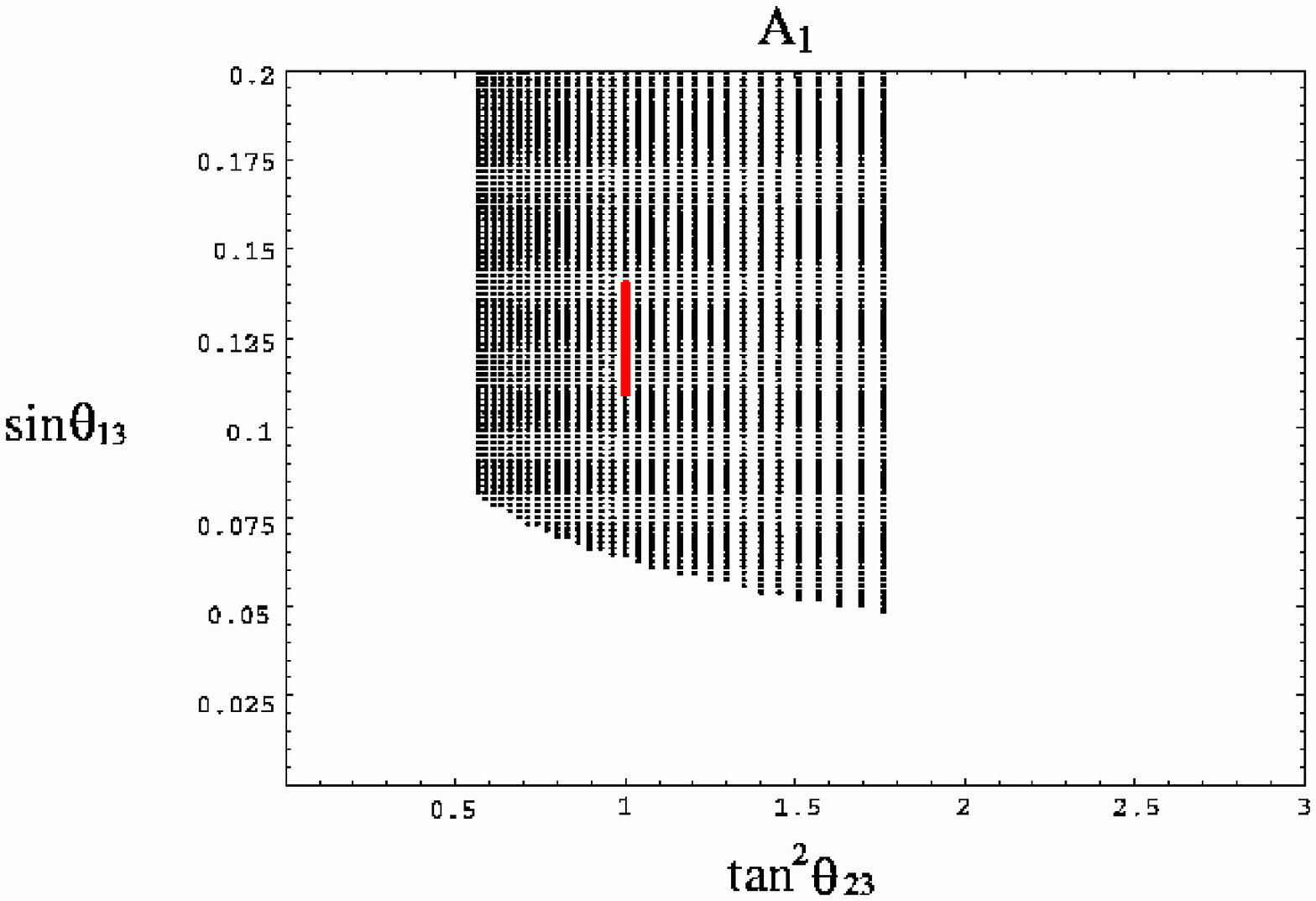}}
\caption{ Scatter plot of $\sin\theta_{13}$ versus $\tan^2\theta_{23}$
for  $A_1$.  The  best fit is shown by a red line.}
\end{figure}
\begin{figure}
\epsfxsize=14.0 cm
\centerline{\epsfbox{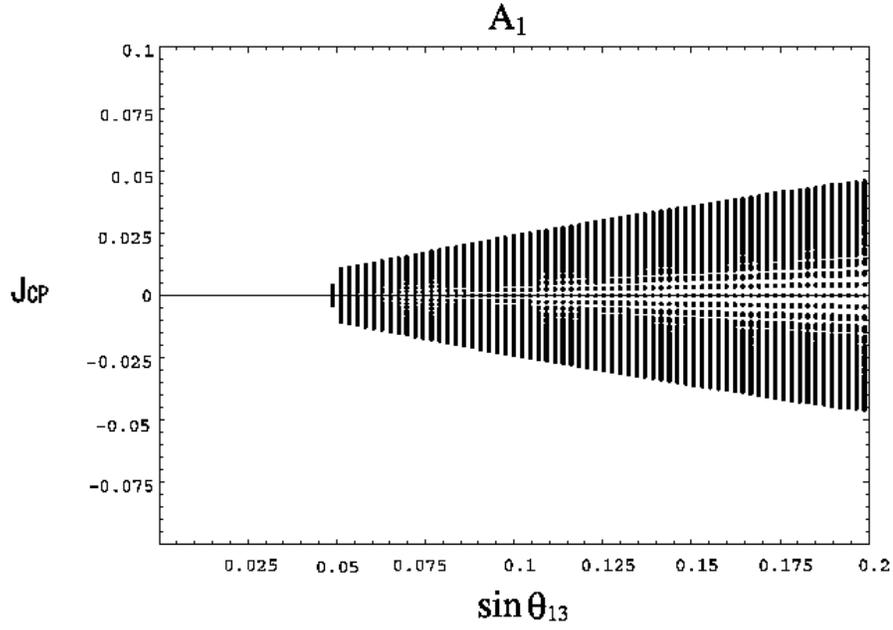}}
\caption{ Scatter plot of $J_{CP}$  versus $\sin\theta_{13}$ for 
$A_1$.}
\end{figure}
\begin{figure}
\epsfxsize=14.0 cm
\centerline{\epsfbox{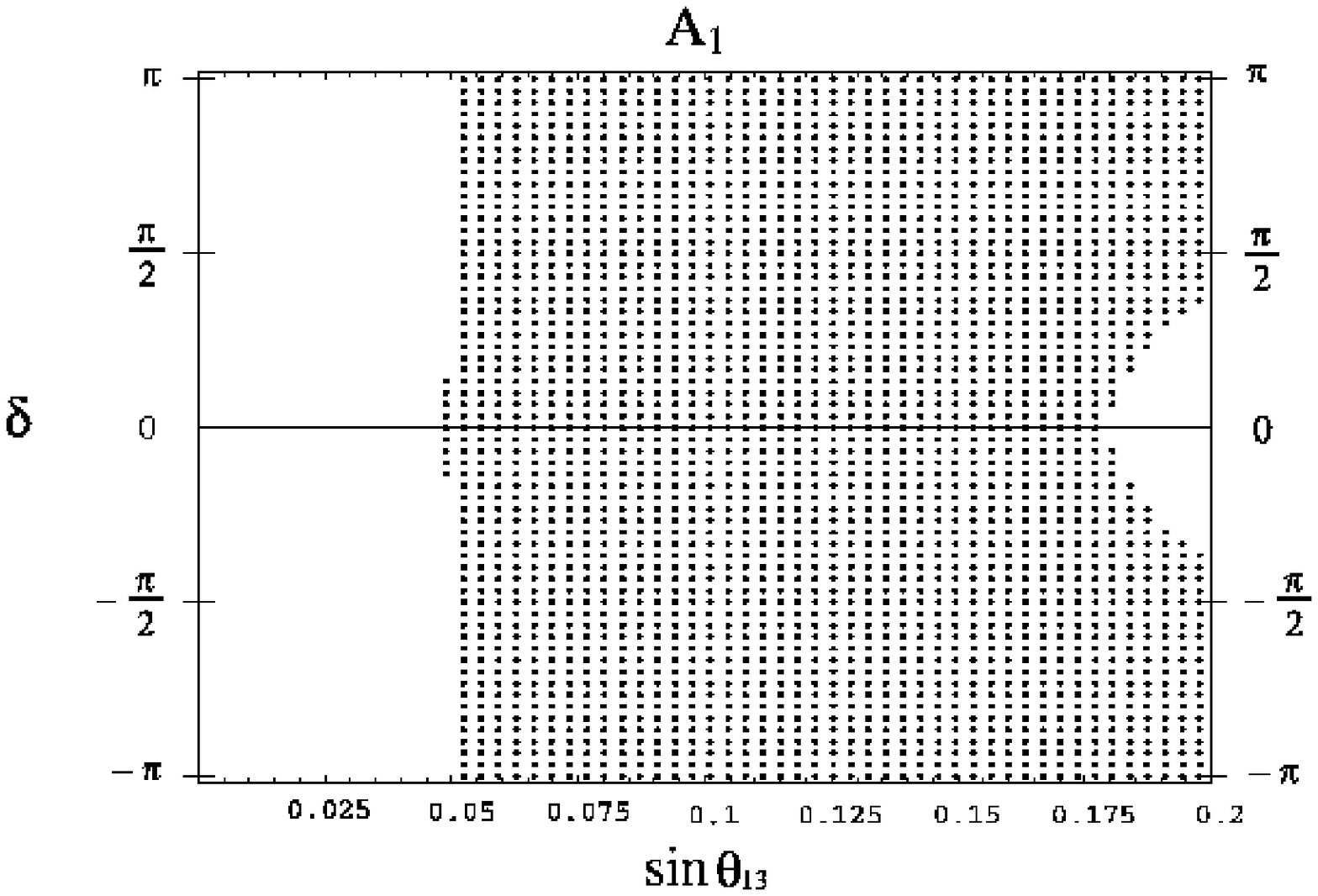}}
\caption{ Scatter plot of $\delta$  versus $\sin\theta_{13}$ for 
$A_1$.}
\end{figure}
\begin{figure}
\epsfxsize=14.0 cm
\centerline{\epsfbox{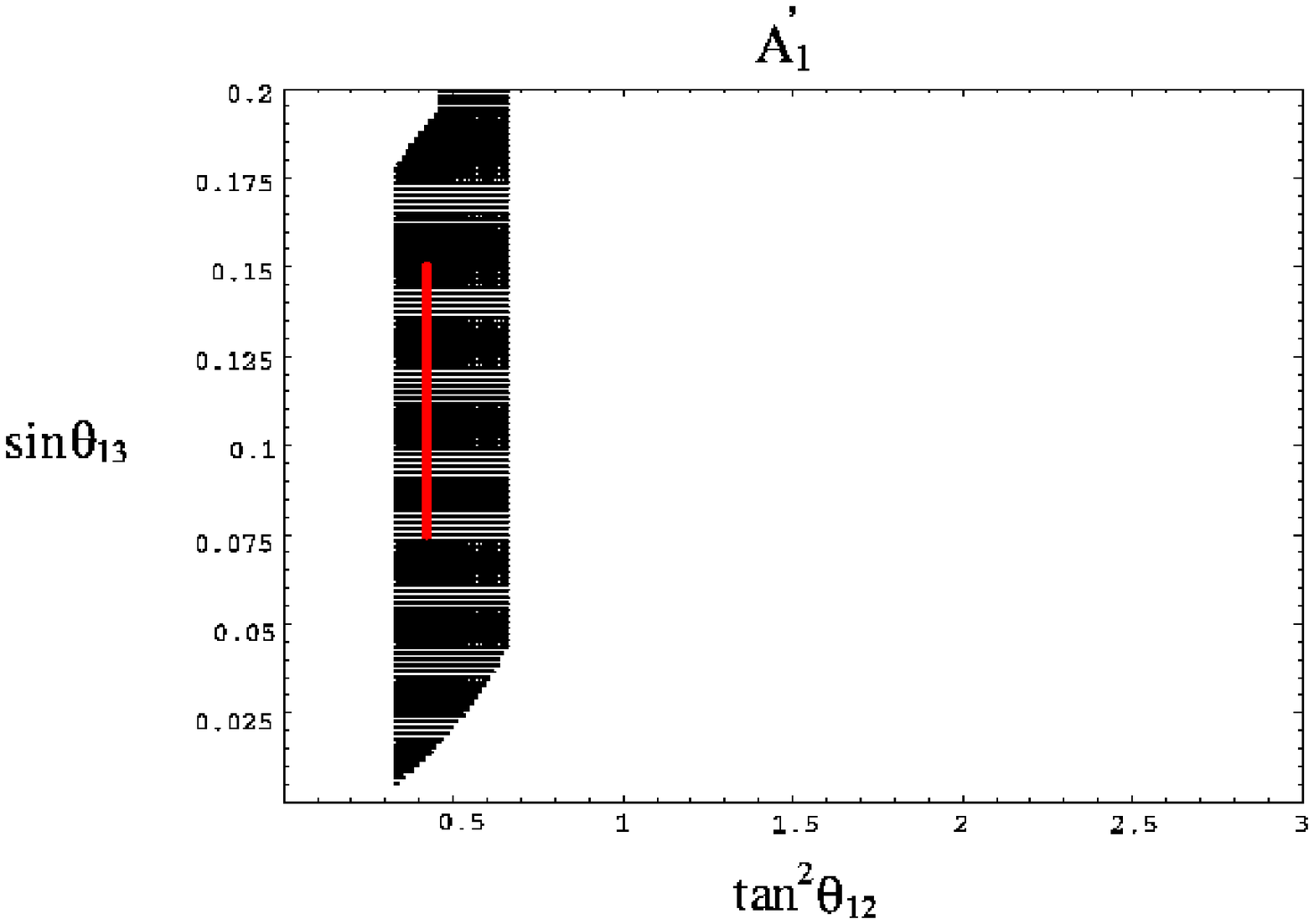}}
\caption{ Scatter plot of $\sin\theta_{13}$ versus $\tan^2\theta_{12}$
  in the case of $\kappa=2\bar\omega=0.07$ for $A_1'$.
 The  best fit is shown by a red line.}
\end{figure}
\begin{figure}
\epsfxsize=14.0 cm
\centerline{\epsfbox{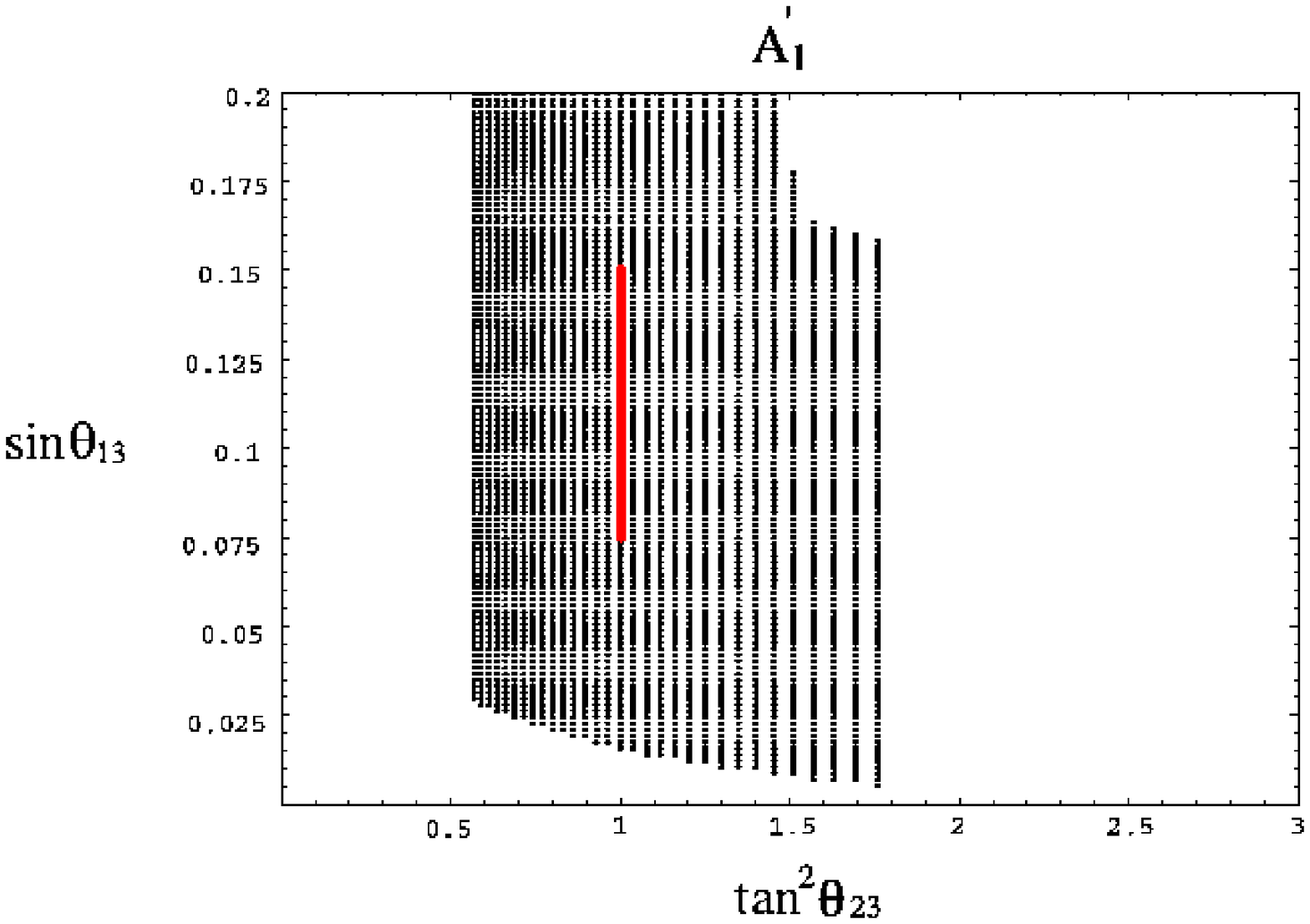}}
\caption{ Scatter plot of $\sin\theta_{13}$ versus $\tan^2\theta_{23}$
 in the case of $\kappa=2\bar\omega=0.07$ for  $A_1'$. 
The  best fit is shown by a red line. }
\end{figure}
\begin{figure}
\epsfxsize=14.0 cm
\centerline{\epsfbox{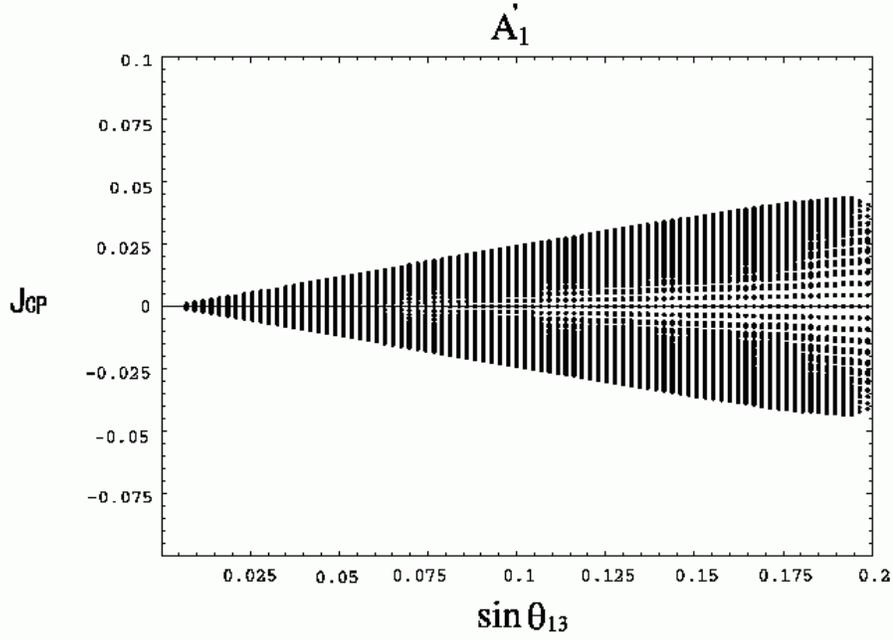}}
\caption{  Scatter plot of $J_{CP}$  versus $\sin\theta_{13}$ 
  in the case of $\kappa=2\bar\omega=0.07$ for  $A_1'$.}
\end{figure}
\begin{figure}
\epsfxsize=14.0 cm
\centerline{\epsfbox{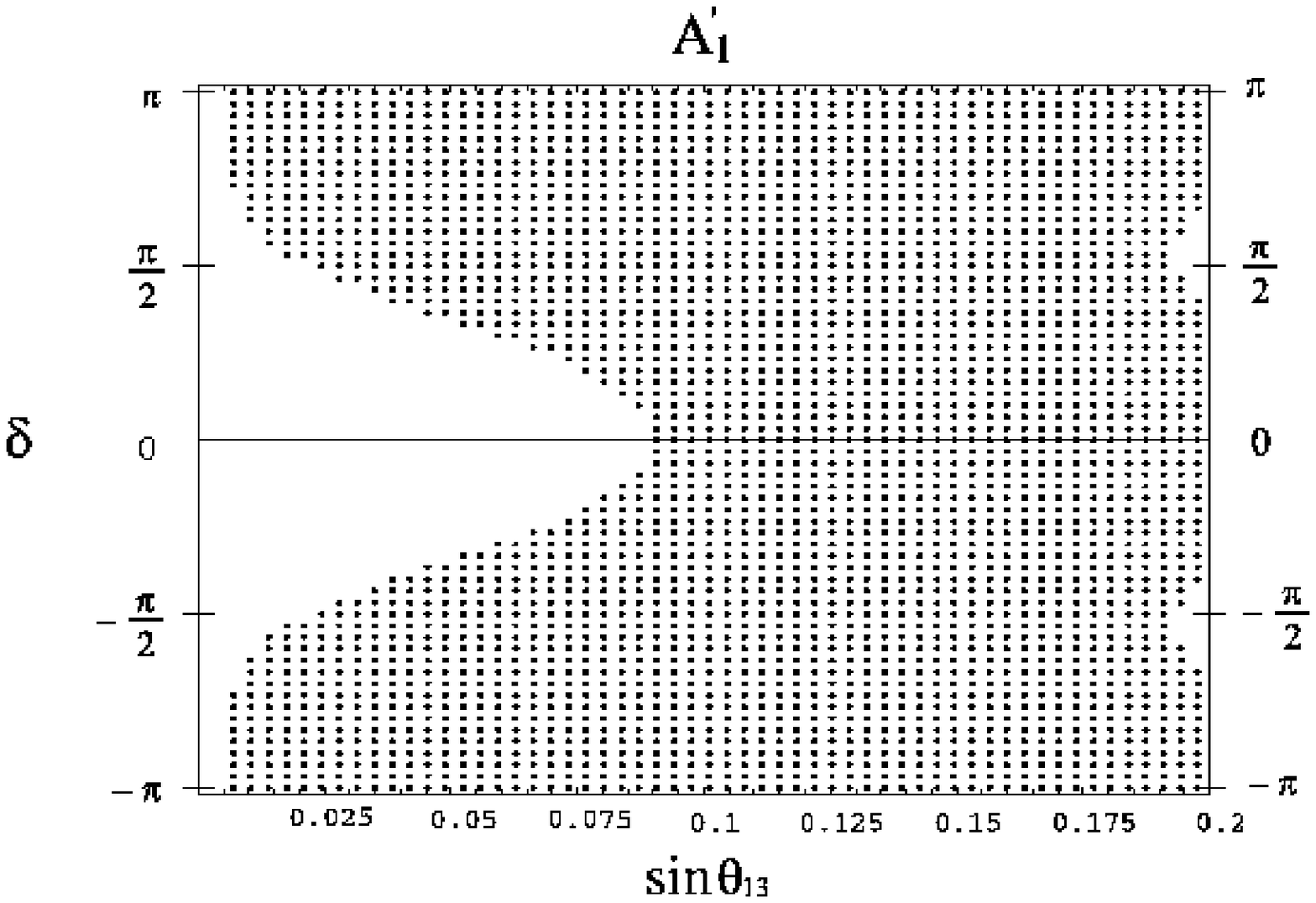}}
\caption{ Scatter plot of $\delta$  versus $\sin\theta_{13}$ for $A_1'$.}
\end{figure}
\begin{figure}
\epsfxsize=14.0 cm
\centerline{\epsfbox{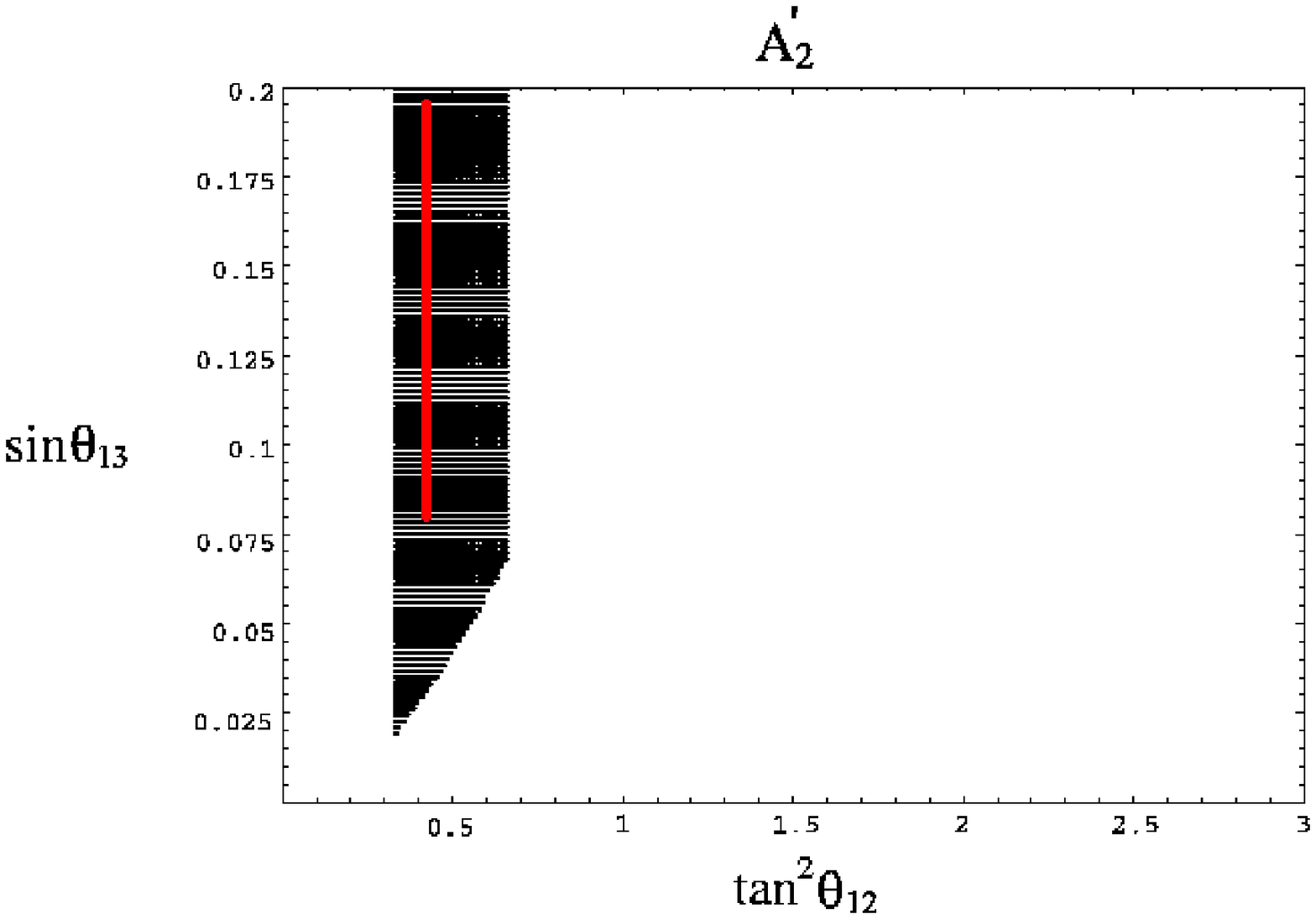}}
\caption{  Scatter plot of $\sin\theta_{13}$ versus $\tan^2\theta_{12}$
 in the case of $\kappa=2\bar\omega=0.07$ for  $A_2'$. 
 The  best fit is shown by a red line.}
\end{figure}
\begin{figure}
\epsfxsize=14.0 cm
\centerline{\epsfbox{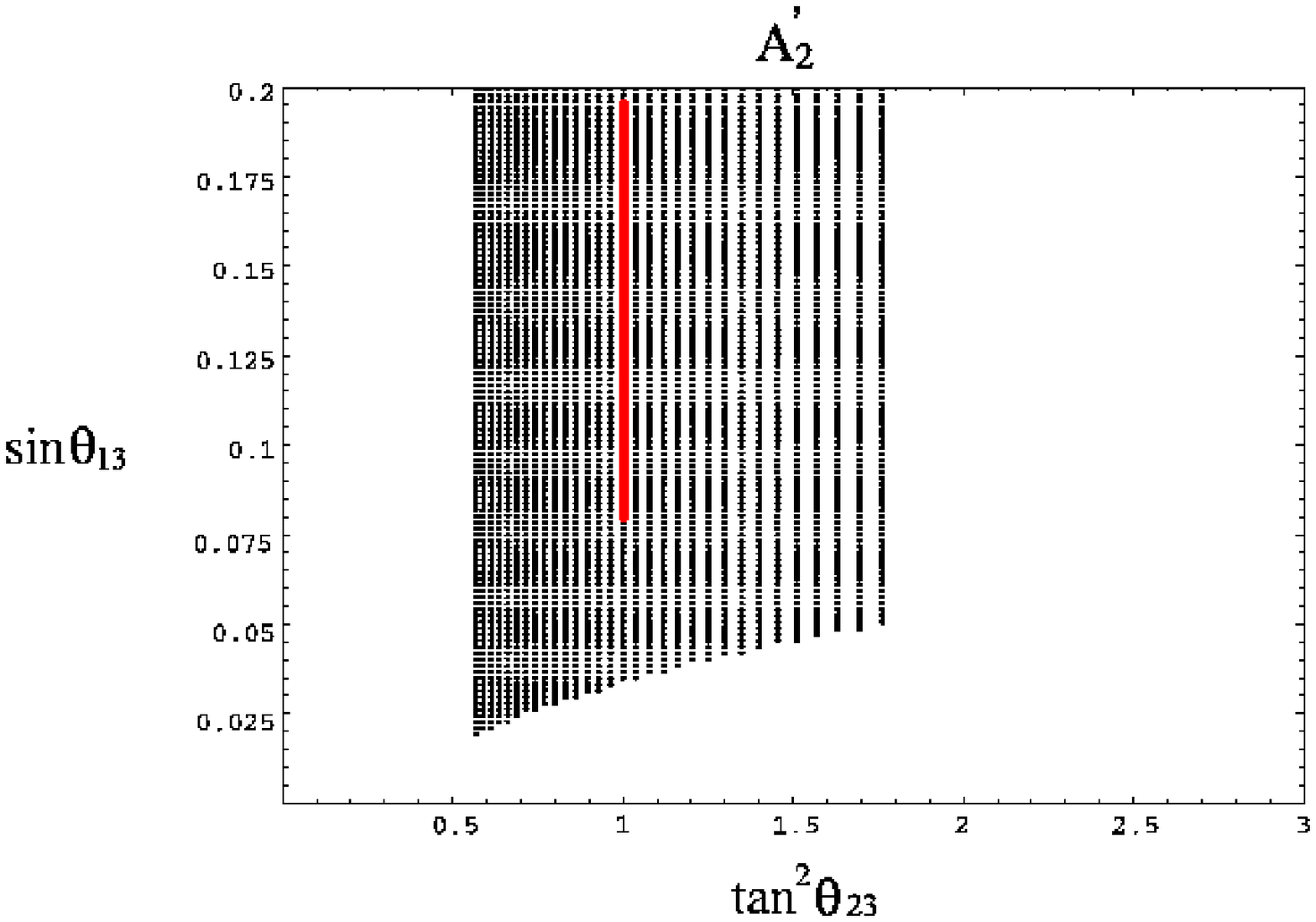}}
\caption{  Scatter plot of $\sin\theta_{13}$ versus $\tan^2\theta_{12}$
  in the case of $\kappa=2\bar\omega=0.07$ for $A_2'$.
 The  best fit is shown by a red line.}
\end{figure}
\begin{figure}
\epsfxsize=14.0 cm
\centerline{\epsfbox{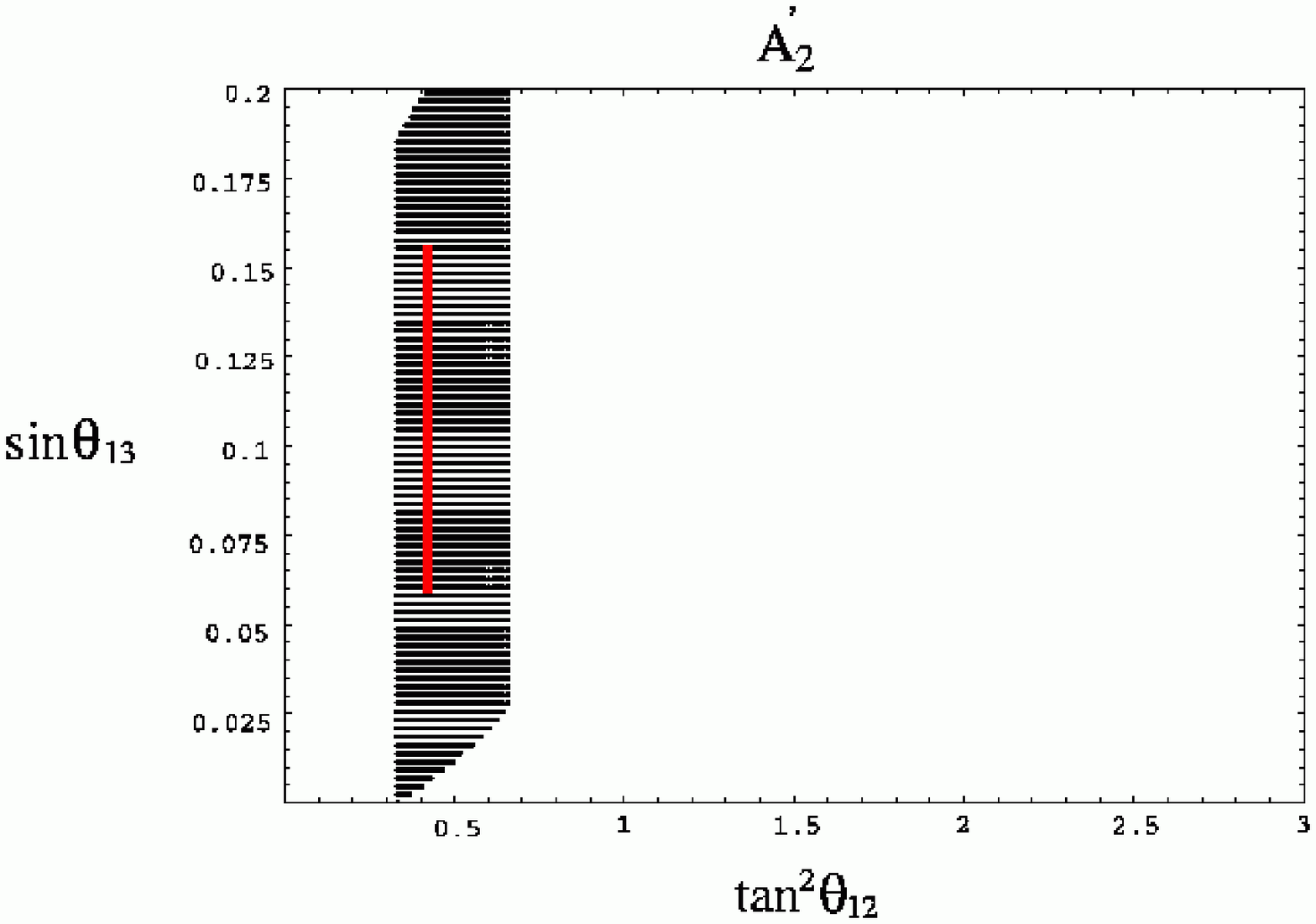}}
\caption{  Scatter plot of $\sin\theta_{13}$ versus $\tan^2\theta_{23}$
in the case of $\tan\theta_{12}\tan\theta_{23}<0$ 
with  $\kappa=2\bar\omega=0.07$ for $A_2'$. 
The  best fit is shown by a red line.}
\end{figure}
\begin{figure}
\epsfxsize=14.0 cm
\centerline{\epsfbox{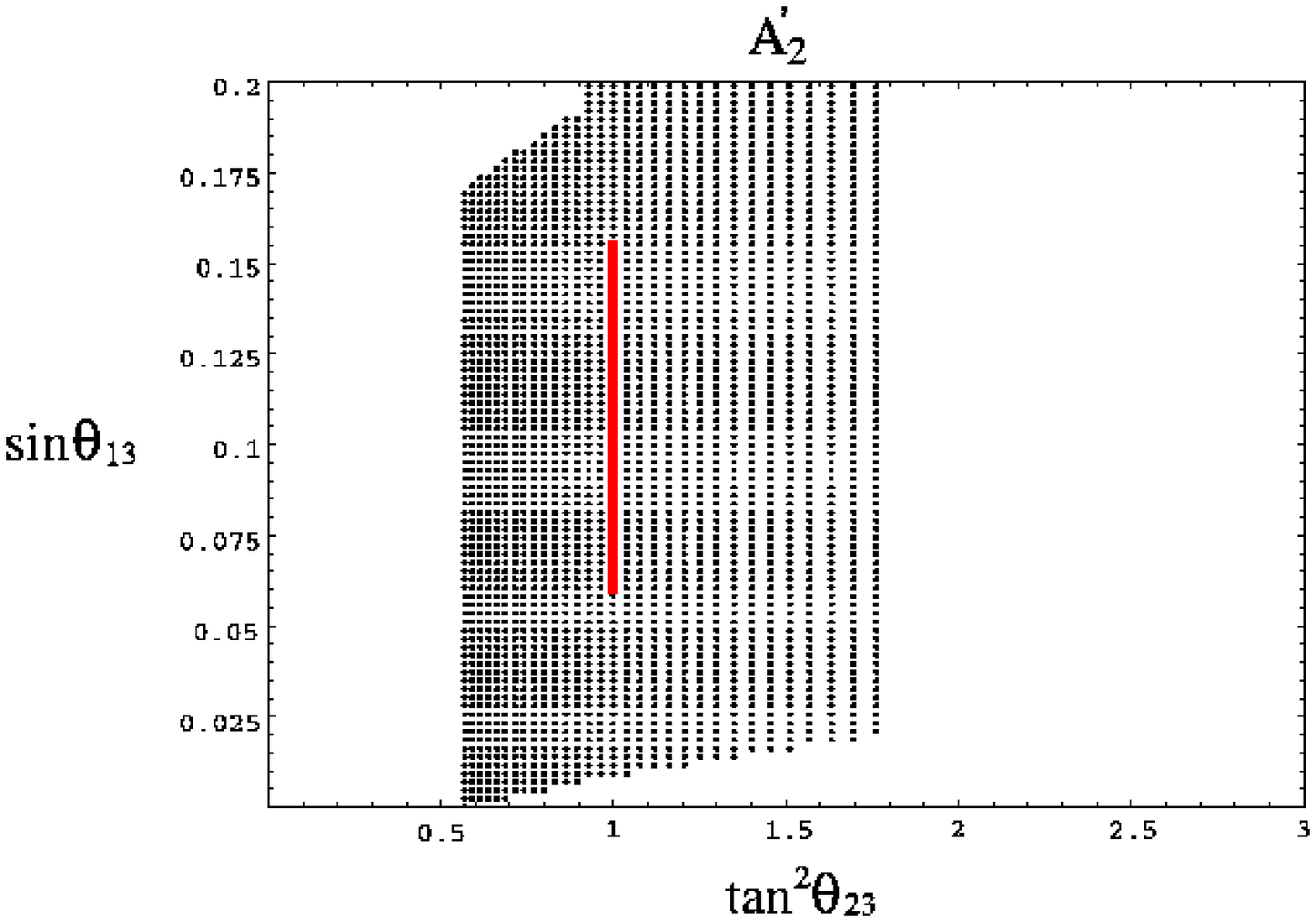}}
\caption{  Scatter plot of $\sin\theta_{13}$ versus $\tan^2\theta_{23}$
   in the case of $\tan\theta_{12}\tan\theta_{23}<0$ 
with  $\kappa=2\bar\omega=0.07$ for $A_2'$.
 The  best fit is shown by a red line.}
\end{figure}
\begin{figure}
\epsfxsize=14.0 cm
\centerline{\epsfbox{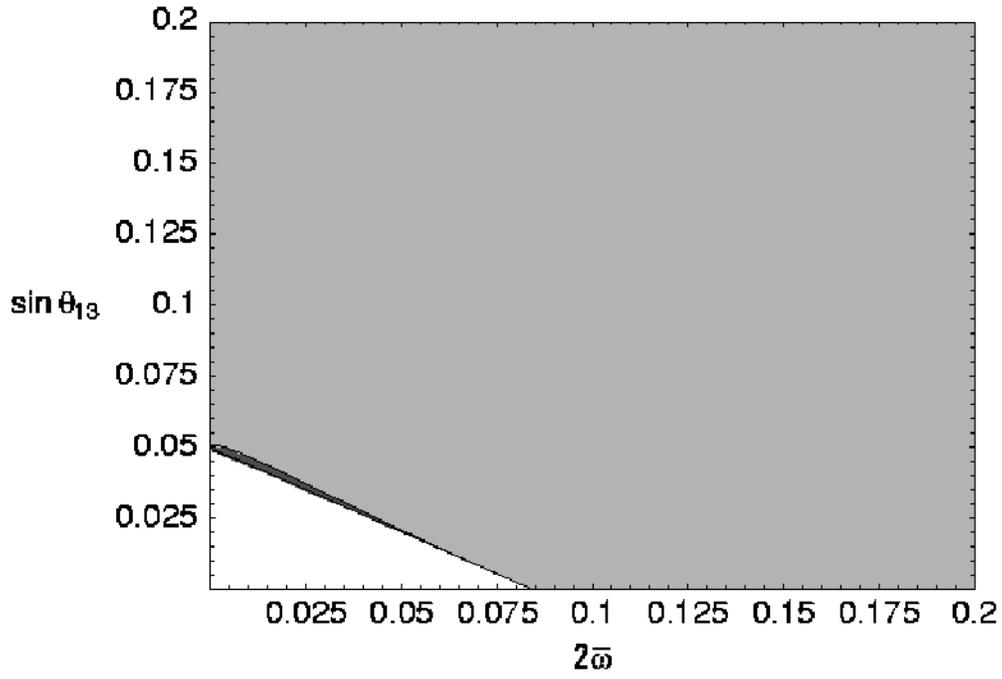}}
\caption{  Prediction of $\sin\theta_{13}$ versus $2\bar\omega$
in the case of $\bar\epsilon=0$.
 The gray region is allowed by the experimental data for  $A_1'$.
  The deep gray region is added  for $A_2'$.}
\end{figure}
\begin{figure}
\epsfxsize=14.0 cm
\centerline{\epsfbox{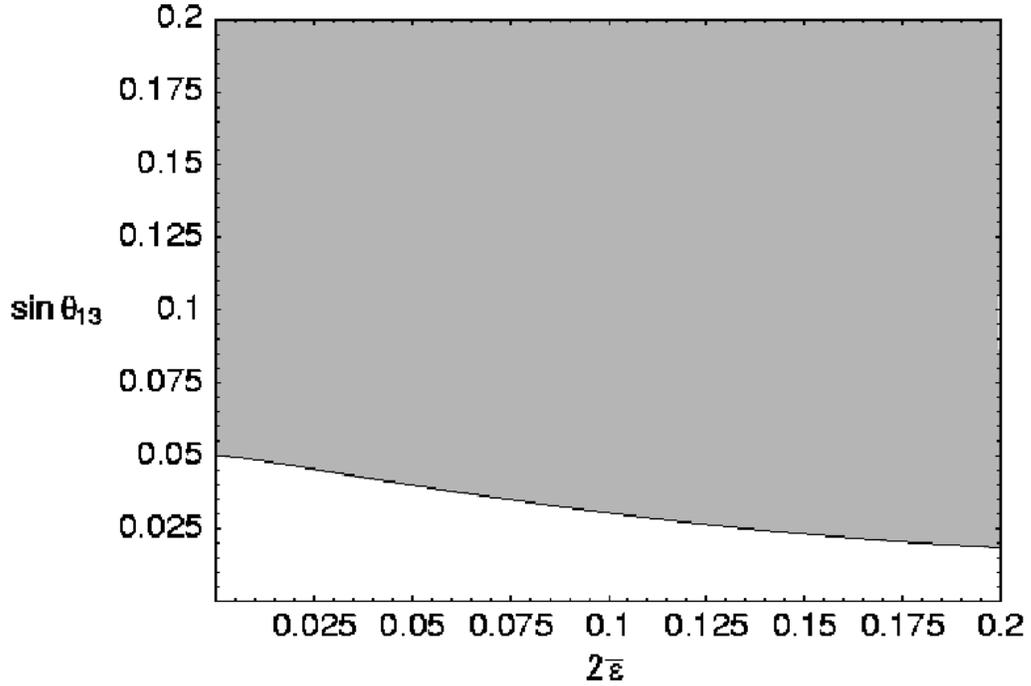}}
\caption{  Prediction of $\sin\theta_{13}$ versus $2\bar\epsilon$
in the case of  $\bar\omega=0$ for both  $A_1'$ and $A_2'$.
 The gray region is allowed by the experimental data.}
\end{figure}
\newpage
\end{document}